\documentclass[usenatbib]{mn2e}
\usepackage{psfig}

\title{Near-IR spectroscopy of PKS1549-79: a proto-quasar revealed?}

\author[M.\,J.\ Bellamy et.\ al.] {M.\,J.\ Bellamy,$^1$\thanks{E-mail:
m.bellamy@sheffield.ac.uk} C.\,N.\ Tadhunter,$^1$ R.\ Morganti,$^2$
K.\,A.\ Wills,$^1$ J.\ Holt,$^1$
\newauthor
M.\,D.\ Taylor,$^1$ and C.\,A.\ Watson.$^1$\\
$^1$Department of Physics and Astronomy, University of Sheffield,
Sheffield S3 7RH, UK \\ $^2$Netherlands Foundation for Research in
Astronomy, Postbus 2, 7990 AA Dwingeloo, The Netherlands\\}

\date{\center{\Large To be submitted for publication in the Monthly
Notices of the Royal Astronomical Society \\
\vspace{.5cm} \today}}

\begin{document}
\maketitle

\begin{abstract}
We present a near-IR spectrum of the nearby radio galaxy
PKS1549-79 (z = 0.153). These data were taken with the aim of
testing the idea that this object contains a quasar nucleus that is
moderately extinguished, despite evidence that its radio jet
points close to our line-of-sight.
We detect broad Pa$\alpha $ emission (FWHM 1745 $\pm
$ 40 km s$^{-1}$), relatively bright continuum emission, and a
continuum slope consistent with a reddened quasar spectrum (3.1 $<$
A$_{V}$ $<$ 7.3), all emitted by
an unresolved point source. Therefore we conclude that we have,
indeed, detected a hidden quasar nucleus in PKS1549-79. Combined with
previous results, these
observations are consistent with the idea that PKS1549-79 is a young
radio source in which the cocoon of debris left over from the
triggering events has not yet been swept aside by circumnuclear outflows.
\end{abstract}


\section{INTRODUCTION}

In recent years, much AGN research has been
conducted in the context of {}``unified schemes'' (e.g. \citealt{barthel89});
i.e. that differences between certain types of active galaxy can
be explained by orientation effects. A great deal of AGN
research is based around refining and constraining such
models. However, the simplest unified models generally describe a static and
unchanging situation, a steady-state regime, and this is unlikely to
be realistic. As the activity evolves it
is likely to affect the distribution of ISM surrounding the nucleus,
potentially influencing the evolution of the host galaxy. To better
understand these processes it is
necessary to develop a more dynamic, evolutionary model for AGN, and to
this end it is important to study sources at different evolutionary
stages.

The existence of compact steep-spectrum (CSS) and gigahertz-peaked
(GPS) radio sources, which are intrinsically small, illustrates the fact
that there is a large range in
the physical dimensions of extragalactic radio sources. The
compactness of these
sources must either be due to frustration, where a dense ISM is
inhibiting the expansion of the radio lobes, or to youth, in which
case the radio lobes simply have not yet had time to expand. Of these
two possibilities youth seems the most likely
(e.g. \citealt{fanti95}). Broad forbidden lines are also much more common
in these compact sources,
suggesting that the cores of these sources are more unsettled
(\citealt{gelderman94}). Moreover, \citet{baker02} find an anti-correlation
between C \textsc{iv}
absorption - which they show to be associated with dusty regions - and radio
source dimensions; the absorption becoming less pronounced with
increasing size. It is therefore
becoming apparent, and perhaps expectedly so, that young radio-loud AGN suffer
greater dust extinction than their more evolved
counterparts, with the nucleus still surrounded by a cocoon of dust
and gas.

Based on its emission line widths and kinematics, the southern radio galaxy
PKS1549-79 (z = 0.153) appears to be just such a young
source. The observed properties of this unusual object are discussed
in detail in \citet{tadhunter01} (hereafter T2001). The flat spectrum,
compactness of the radio emission ($\sim$150 milliarcsec, or $\sim$540
pc\footnote{We assume cosmological
parameters of H$_{0}$ = 50 km s$^{-1}$ Mpc$^{-1}$ and
q$_{0}$ = 0 throughout.}) and one-sided jet morphology of this source
indicate that its radio axis is aligned close to our line-of-sight. In the
standard unification model, the source would be expected to resemble a
quasar at optical wavelengths, with non-stellar continuum emission and
broad permitted lines. However, optical spectra of
PKS1549-79 show a predominantly stellar
continuum with no good evidence for broad permitted lines, but with
strong and
unusually broad forbidden lines ($\sim $1300 km s$^{-1}$). Amongst
powerful radio galaxies such broad forbidden lines are only ever
detected in sources that are intrinsically compact. The blueshift
of the [O
\textsc{iii}]$\lambda $$\lambda $5007, 4959 emission lines relative to
low-ionisation forbidden lines and the H\textsc{i} 21 cm absorption
line suggests that the [O \textsc{iii}] emitting region is outflowing at 600
$\pm $ 60 km s$^{-1}$ (T2001, \citealt{morganti01}).

Overall, observations suggest that there is significant dust
obscuration in this object even though our line-of-sight is
close to the radio axis. This is entirely consistent with PKS1549-79
being a young radio source in which the obscuring material around the
nucleus has not yet been swept aside (T2001). If this model is
correct, near-infrared observations - being subject to only $\sim$10\%
of the visual extinction - should show a bright, non-stellar continuum
and broad permitted lines as the quasar shines through the ISM. This is the
prediction we test in this paper.

\section{DATA COLLECTION AND REDUCTION}

The K-band infrared spectra were taken in shared risks service mode on
26 July 2002 using the IRIS2 instrument on the AAT. The
Sapphire 240 grism was used with a 1 arcsec slit, oriented North-South
on the sky. The seeing was reported to be 1.1 arcsec throughout the
observations; analysis of stars in the 10 second exposure  acquisition
image is in good
agreement with this (see Section 3.1). 22 exposures of 300 seconds
were obtained for PKS1549-79; the
galaxy was `nodded' between two positions on the slit in an ABBA pattern
with 11 exposures at each position. Four exposures of 80 seconds were
taken of the A0 star HIP77712 at a similar airmass for calibration
purposes.

Each set of 11 galaxy exposures was co-added in the \textsc{iraf}
package using median filtering to remove cosmic rays; the median `B'
image was then subtracted from the median `A' image to remove the
night sky lines. The standard star frames were combined in a similar
way. The resulting galaxy and star frames were then flat-fielded using a
dark-subtracted flat field taken with the same instrumental setup.

A xenon arc frame was used to make a two-dimensional wavelength
calibration which was then applied to the data frames. The original
spectra cover the range 20109 - 24640 \AA~ but the useful data
presented here are over the range 20500 - 23000 \AA. From the
night sky lines, the accuracy of the wavelength calibration was found
to be $\pm $ 0.65 \AA~ ($\pm $ 9 km s$^{-1}$) and the systematic
errors are estimated to be $<$ 1 \AA~ ($<$ 14 km
s$^{-1}$). Fits to the night sky lines show the spectral
resolution to be 10.1 $\pm $ 0.4 \AA; the pixel scale of 4.43 \AA~
pix$^{-1}$ is adequate to sample this. The spatial pixel scale of the
acquisition image is 0.446 arcsec pix$^{-1}$. One-dimensional spectra were
extracted from the two-dimensional frames, each from 11 pixel rows to
aid accurate flux calibration.

The exposures of HIP77712 were used to flux calibrate the data, with
the assumption that the intrinsic spectral energy distribution (SED)
of the star is that of a perfect
black body at T = 9480 K. The magnitude-to-flux conversion was
performed with reference to \citet*{bessell98}.

\begin{figure}

\centerline{\psfig{figure=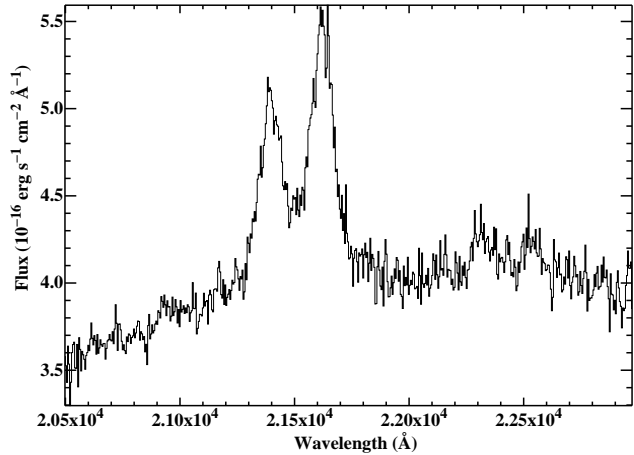,width=9cm,angle=0.}}

\caption{K-band spectrum of PKS1549-79 obtained with the IRIS2
instrument on the
AAT.  Doubling of the lines is caused by birefringence associated with
the grism used for the observations (See Section 2).}

\label{fig: data}

\end{figure}

During data reduction it became apparent that there was a problem with
the IRIS2 instrument in spectroscopy mode at the time of the
observations. All the spectra we had
obtained, including arc calibration frames, showed a splitting of
spectral features in the wavelength direction. After consulting with the AAT
team we were informed that the manufacturers had mounted
the Sapphire grisms at right-angles to their intended
orientation and that a birefringence effect was leading to the
observed doubling.  The spectra we obtained are composites of the
actual spectrum and a copy of this spectrum, blueshifted
by around 50 pixels ($\sim$230 \AA~ in the case of these data) relative
to the original and at a slightly lower intensity. However, the
splitting caused by the birefringence is large enough that useful
kinematic information can be extracted for individual lines in the
spectra. The final, reduced K-band spectrum of PKS1549-79 is shown in
Fig. \ref{fig: data}. The Starlink \textsc{figaro} and \textsc{dipso}
packages were used to analyse the data.

\section{RESULTS}

\subsection{Paschen alpha}

\begin{figure}

\centerline{\psfig{figure=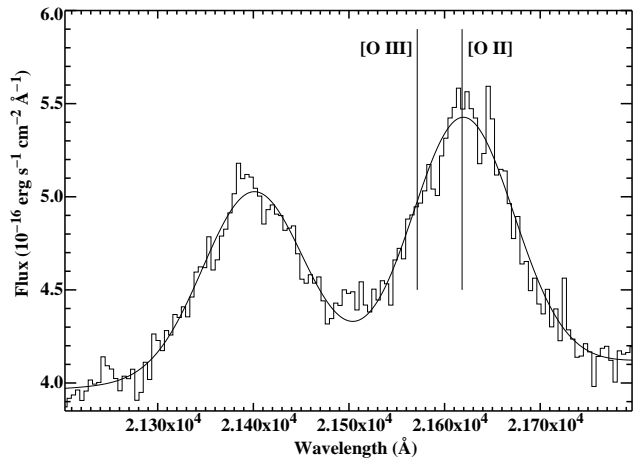,width=9cm,angle=0.}}

\caption{\textsc{dipso} fit to the Pa$\alpha $ line(s). Pa$\alpha $
wavelengths corresponding to the [O \textsc{ii}] and [O \textsc{iii}]
redshifts (T2001) are marked for comparison.}

\label{fig: pafit}

\end{figure}

\begin{table*}
\begin{center}\begin{tabular}{lcc}
\hline  Parameter& Pa$\alpha $& H$_{2}$ or [Si \textsc{vi}]\\
\hline
\hline  FWHM (km s$^{-1}$)& 1745 $\pm $ 40& 1100 $\pm $ 180\\
z&0.15266 $\pm $ 0.00007& 0.1525 $\pm $ 0.0002 or 0.1494 $\pm $ 0.0002\\
Total line flux (erg cm$^{-2}$ s$^{-1}$)& (3.16 $\pm $ 0.08) $\times $ 10$^{-14}$& (3.5 $\pm $ 1.0)
$\times $ 10$^{-15}$\\
Continuum flux (erg cm$^{-2}$ s$^{-1}$ \AA$^{-1}$)& (4.0 $\pm $ 0.1)
$\times $ 10$^{-16}$& (4.0 $\pm $ 0.2) $\times $10$^{-16}$\\
Equivalent width (\AA)& 79 $\pm $ 3& 9 $\pm $ 3\\
Rest-frame equivalent width (\AA)& 69 $\pm $ 3& 8 $\pm $ 3\\ \hline

\end{tabular}\end{center}

\caption{Results of the \textsc{dipso} fits to the spectral lines. The
instrumental width has a negligible effect on the measured FWHMs.}
\label{table: results}
\end{table*}
A
single Gaussian provides an adequate fit to each of the two components
of the split line.
The best fit to the Pa$\alpha $ line(s) is
shown in Fig. \ref{fig: pafit}, with the redder peak being at the correct
wavelength. A constraint was applied that the two Gaussians should
have the same
width. The results of the fitting are summarised in Table \ref{table: results}.

The fit gives a FWHM of 126 $\pm $ 3 \AA~ (1745 $\pm $ 40 km
s$^{-1}$ in the rest frame). This value is consistent with the widths
of permitted lines in broad line radio galaxies
(BLRG) and quasars (\citealt*{hill96}, \citealt{marziani01}),
although it is at the lower end of the range. The relatively small
broad-line widths are also consistent with the radio jet being
oriented close to our line-of-sight (\citealt{wills86}).

The redshift of the Pa$\alpha $ line is found to be z = 0.15266 $\pm $
0.00007. This is consistent with the {[}O \textsc{ii}{]} redshift of z = 0.1526
$\pm $ 0.0002 but inconsistent with the {[}O \textsc{iii}{]} (outflow) redshift
of z = 0.1501 $\pm $ 0.0002 (T2001). This indicates that the Pa$\alpha $
emission originates in a broad line region (BLR) at the systemic
redshift. The Pa$\alpha $ wavelengths corresponding to the
{[}O \textsc{ii}{]} and {[}O \textsc{iii}{]} redshifts
are marked on \ref{fig: pafit} for comparison.

The K-band continuum flux level is $\sim$4$\times $ that measured in the
optical spectrum of T2001, despite the fact that the optical spectrum
was obtained using a larger aperture (4.3 by 5 arcsec). This is
inconsistent with the SED of an unreddened stellar
population, particularly a young population,
but is consistent with that of a moderately obscured quasar source
shining through the intervening ISM. A power-law was fit to the
continuum to aid
comparisons with quasar continua (see Section 4). The spectral index
of this power-law is $\alpha $ =
2.3 $\pm $ 0.1 (F$_{\upsilon }$ $\propto $ $\nu $$^{-\alpha }$, F$_{\lambda }$
$\propto $ $\lambda $$^{\alpha -2}$). This index is significantly
larger than that measured for the 16 unreddened sources in the
\citet{simpson00} sample of radio-loud quasars (-0.67 $<$ $\alpha $ $<$
1.62), but similar to the
two reddened quasars in that sample. If
the intrinsic spectrum of PKS1549-79 is quasar-like then 
this indicates significant reddening. 

The combined flux of both Pa$\alpha $ components is (3.16 $\pm $ 0.08)
$\times $
10$^{-14}$ erg cm$^{-2}$ s$^{-1}$. This
flux gives a rest-frame equivalent width of
69 $\pm $ 3 \AA~ for an assumed continuum flux of (4.0 $\pm $
0.1)
$\times $ 10$^{-16}$ erg cm$^{-2}$ s$^{-1}$ \AA$^{-1}$ in the observed
frame. For
comparison, the rest-frame
Pa$\alpha $ equivalent widths measured in the low-redshift quasars 3C
273 and PDS 456 are 167 $\pm $ 16 \AA~ and 120 $\pm $ 8
\AA~respectively (\citealt{hill96}, \citealt{simpson99}). Both of these
quasar values are
significantly larger than that measured in PKS1549-79. This could
indicate the presence of a beamed, non-thermal continuum component in
PKS1549-79  which would further support the theory
that the radio jet is oriented towards our line-of-sight.

If there is a quasar nucleus shining through at near-IR
wavelengths then the emission from the galaxy should be dominated by a
central, unresolved source. To test this, the spatial intensity profile of
PKS1549-79 was compared with those of four bright stars in the
acquisition image. The mean FWHM of the stellar 2-D Gaussian measures 1.13 $\pm
$ 0.06 arcsec in the x-direction by 1.23 $\pm $ 0.06 arcsec in the
y-direction. In comparison, the profile of PKS1549-79 measures 1.13
$\pm $ 0.17 by 1.43 $\pm $ 0.15 arcsec. Therefore PKS1549-79
appears quasar-like at near-IR wavelengths.

\subsection{Other features}

\begin{figure}

\centerline{\psfig{figure=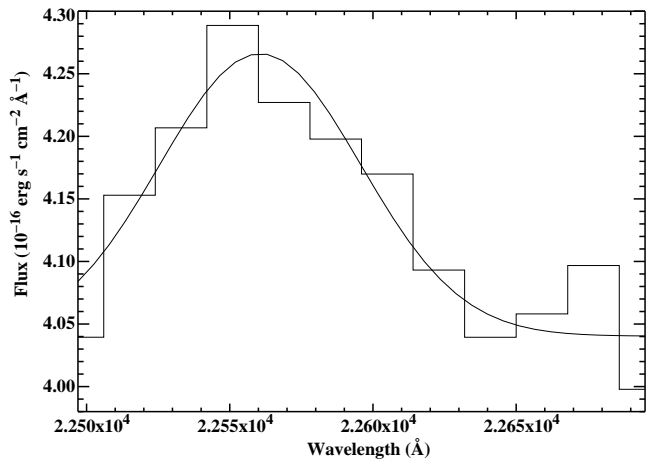,width=9cm,angle=0.}}

\caption{\textsc{dipso} fit to the $\sim$22550 \AA~ line.}

\label{fig: hfit}

\end{figure}

Apart from the Pa$\alpha $ line, the only other feature in the
spectrum is a second line visible at $\sim$22550 \AA, subject to
the same splitting as the rest of the spectrum. As the line appears
faint, the spectrum was re-binned to one-quarter resolution over
this region to improve the signal to noise. The line is broad
enough for this to be acceptable. The blueshifted split component
is too noisy for an adequate fit to be made and so only the redmost
component was fit in \textsc{dipso}. The fit is shown in
Fig. \ref{fig: hfit} and 
the results of the fitting are given in Table \ref{table: results}.

The FWHM of this line is 83 $\pm $ 14 \AA~(1100 $\pm $ 180 km
s$^{-1}$ in the rest frame). Correcting for the fact that only one
component was fit, the
total flux of the line is (3.5 $\pm $ 1.0) $\times $ 10$^{-15}$ erg
cm$^{-2}$ s$^{-1}$. This corresponds to an isotropic luminosity of
(4.1 $\pm $ 1.2) $\times $ 10$^{41}$ erg s$^{-1}$.

The most likely candidate for this emission is the H$_{2}$ $\nu $ = 1-0
S(3) line, our identification is based on redshift
arguments. The H$_{2}$ $\nu $ = 1-0 line would lie at the same
redshift as the Pa$\alpha $ line,
whereas other lines imply large inflow or outflow velocities. However,
[Si \textsc{vi}] is also a possibility since the implied outflow velocity
and width 
are broadly consistent with the [O \textsc{iii}] outflow parameters
(T2001). The possibility that this feature is a blend of both lines
cannot be discounted. However, the low S/N of the data - coupled with
the uncertainties introduced by the splitting of the spectrum - means that
no definite conclusions can be drawn.

If the feature is indeed H$_{2}$ $\nu $ = 1-0 S(3), then PKS1549-79 is one
of the most luminous molecular hydrogen emitters in the local
universe, surpassing the H$_{2}$ luminosity of Cygnus A by almost an
order of magnitude (\citealt{ward91}, \citealt*{tsr99}).

Unfortunately no other molecular hydrogen lines are visible in the
spectrum with which to compare this result. This is to be expected
however, as the H$_{2}$ $\nu $ = 1-0 S(3)/[Si \textsc{vi}] feature is
only weakly detected and the only molecular line expected to be
comparable in luminosity (H$_{2}$ $\nu $ = 1-0 S(1)) is redshifted out
of the useful range of our data.

%




\section{DISCUSSION}

Our observations provide clear evidence that
emission from the central regions of PKS1549-79 is significantly
affected by dust obscuration. In this section we attempt to estimate
the degree of reddening\footnote{We assume the standard interstellar
extinction law of \citet{whitford58}.} and correct for it in order
to obtain a
more realistic estimate of the intrinsic Pa$\alpha $ luminosity. For
comparison, T2001 derive a visual extinction in the
range 0.23 $<$
A$_{V}$ $<$ 18 mag, based on H\textsc{i} 21 cm absorption
characteristics.

Firstly, we try to determine the reddening directly from Balmer line
ratios, these being relatively insensitive to variations in the
physical conditions of the
emitting medium. With reference to \citet{tadhunter93}, the
Pa$\alpha $/H$\beta $ line ratio is found to be 25 $\pm $ 1. Assuming
Case B recombination, a temperature of 10,000K and an electron
density of 10$^{4}$ electrons cm$^{-3}$, this line ratio gives an
E$_{B-V}$ = 1.29 $\pm $ 0.01 (\citealt{seaton79}, \citealt{osterbrock89}),
corresponding to a visual extinction of A$_{V}$ = 4.00 $\pm $
0.13 mag (A$_{V}$ = R $\times $ E$_{B-V}$, assuming R = 3.1 $\pm $
0.1). However, there is a great deal of uncertainty associated with
this value because the H$\beta $ emission is likely to
be dominated by a
narrow line region (NLR) component. The BLR emission from within the
obscuring region may
then contribute only a fraction, if any, of the total H$\beta $ flux,
and therefore the
derived extinction is likely to be an underestimate. Recently obtained
high-resolution spectra of PKS1549-79 include
H$\alpha $ emission: however, the H$\alpha $ line is contaminated with
bright N \textsc{ii} recombination lines and an atmospheric absorption
line. These features will significantly reduce the accuracy of line
fittings to the H$\alpha $ line and for this reason the line has not
been used in this analysis.

Another approach is to fit a power-law to the near-IR continuum and then
calculate the de-reddening necessary to make the slope consistent with
a quasar spectrum. As stated in Section 3.1, the spectral index of the
power-law fit to our data is $\alpha $ = 2.3 $\pm $ 0.1. Typical
unreddened quasar spectral
indices lie in the range -0.67 $<$ $\alpha $ $<$ 1.62 and average
$\bar{\alpha}$ = 0.90 (\citealt{simpson00}). Clearly the spectrum for
PKS1549-79 is considerably redder than even the upper
limit. De-reddening of the spectrum to match the quasar range of
values requires a
visual extinction in the range 1.8 $<$ A$_{V}$ $<$ 8.0 with A$_{V}$
= 3.8 for the average case. The full de-reddening results are given in
Table \ref{table: reddening}. If the spectrum is contaminated by a
beamed, non-thermal component then the intrinsic continuum shape may
not resemble a quasar; in the extreme case it may more closely
resemble a BL Lac. The 6 BL Lacs in the \citet{massaro95} sample show
a tighter range of spectral indices (0.86 $<$ $\alpha $ $<$ 1.19) with
an average that is redder than that of the quasar sample: this may
imply a smaller degree of reddening. However, as the full range of BL
Lac spectral indices lies within the broader range of quasar spectral
indices, this result has no significant effect on our analysis.

\begin{table*}
\begin{center}\begin{tabular}{cccc}
\hline  Reddening Degree   $\longrightarrow$& Minimum& Average&
Maximum\\ \hline \hline
Spectral Index $\alpha $& 1.62& 0.90& -0.67\\
E$_{B-V}$& 0.60& 1.23& 2.62\\ A$_{V}$ (total)& 1.8& 3.8& 8.0\\
A$_{V}$ (intrinsic)$^{a}$& 1.2& 3.1& 7.3\\
Pa$\alpha $ Flux
(10$^{-14}$ erg s$^{-1}$ \AA$^{-1})$& 3.95 $\pm $ 0.10& 5.02 $\pm $
0.13& 8.50 $\pm $ 0.22\\ Pa$\alpha $ Luminosity$^{b}$ (10$^{41}$ erg s$^{-1})$&
46.4 $\pm $ 1.2& 59.0 $\pm $ 1.5& 99.8 $\pm $ 2.5\\ \hline
\end{tabular}\end{center}

\caption{Resulting Pa$\alpha $ emission parameters corresponding to
the different degrees
of reddening as discussed in the text.\newline$^{a}$Assuming Galactic
extinction A$_{V}$ = 0.68
(\citealt{schlegel98}).\newline$^{b}$Calculated assuming
isotropic emission.}
\label{table: reddening}
\end{table*}
To investigate the contribution such reddened power-law continua would
make at optical wavelengths, curves were generated for the derived
power-laws and then re-reddened by the corresponding
amount. Fig. \ref{fig: red}
shows the results of this for the maximum, average and minimum
extinctions/power-laws derived above. Certainly, the minimum reddening
case is ruled out on the basis that the quasar alone would
contribute more flux at $>$6000 \AA~ than is detected (T2001). The
average extinction, corresponding closely with that derived from the
Pa$\alpha $/H$\beta $ ratio, would also be contributing around half
the observed flux in this region and also with a marked slope
that one would expect to detect superimposed on the starlight:
modelling shows no evidence for a power-law component in this
region (\citealt{tadhunter02}). This implies that
the actual reddening is closer to the upper limit than to the
average, regardless of whether the intrinsic continuum slope more
closely resembles a quasar or BL Lac. It seems safe to conclude that
the total visual
extinction lies in the range 3.8 $<$ A$_{V}$ $<$ 8.0 mag. This
corresponds to an intrinsic extinction in the host galaxy of 3.1 $<$
A$_{V}$ $<$ 7.3, having corrected for the Galactic extinction of
A$_{V}$ = 0.68 ({\citealt*{schlegel98}).

\begin{figure}

\centerline{\psfig{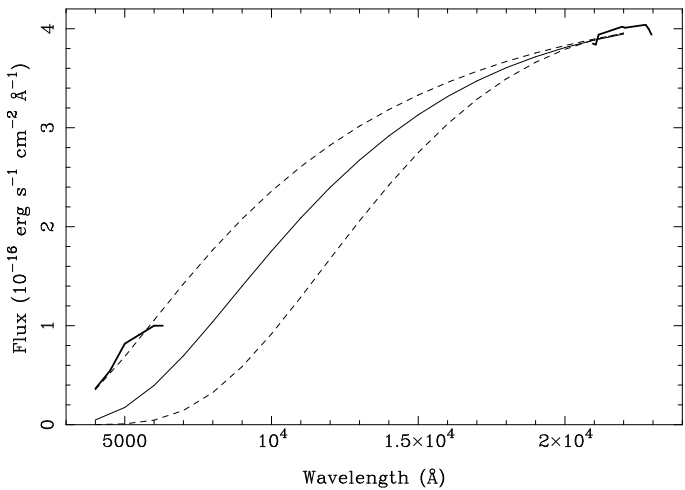}}

\caption{Reddened power-laws generated for the parameters given in
Table \ref{table: reddening}. The average case is shown by the solid line,
the minimum case
by the upper broken line and the maximum case by the lower broken
line. The thick solid lines represent the optical continuum data from
T2001 and the near-IR data from this paper. The data are plotted at the
observed wavelengths and the model shifted into this frame to match.}

\label{fig: red}

\end{figure}

The de-reddened Pa$\alpha $ fluxes for the three cases are
given in Table \ref{table: reddening}, along with the corresponding
isotropic
luminosities. Comparison with similarly derived luminosities for a
sample of quasars and BLRGs
(\citealt{rudy82}, \citealt{hill96} and \citealt{simpson99}) shows that the
luminosity corresponding to the average reddening ((59.0 $\pm $ 1.5)
$\times $ 10$^{41}$ erg s$^{-1}$) is more than an order of magnitude
less than that measured in the low-z quasars, and at the upper end of the
range measured in the BLRGs. These results are consistent with
PKS1549-79 being a low-luminosity quasar.

%




\section{CONCLUSIONS AND FURTHER WORK}

The detection of a broad Pa$\alpha $ line and a
bright, reddened infrared continuum indicates that we have, indeed, uncovered
an obscured BLR and quasar nucleus in PKS1549-79. The calculated intrinsic
luminosity and equivalent width of the Pa$\alpha $ emission, and the
star-like profile of the near-IR nucleus, further
support this conclusion. Although such findings are not unprecedented
(\citealt{djorg91}, \citealt{hill96}), PKS1549-79
is a special case since the observed radio properties lead us to believe
that the AGN is viewed from a direction close to that of the radio
axis and should therefore
exhibit the optical/infrared properties of radio loud quasars; i.e.
no significant obscuration of the BLR and nucleus. The fact that significant
obscuration is seen suggests that PKS1549-79 is a fundamentally different
type of object that does not fit neatly into standard unified
schemes.

It is our suggestion that PKS1549-79 is an AGN in the early stages
of evolution, a proto-quasar. The high obscuration is a transitory
phase that will pass as the gas and dust is dissipated or blown out of the
ionisation cones by the circumnuclear outflows detected at optical
wavelengths (T2001). These infrared data
support this model and tie in well with expectations.

Since a key assumption of the model is that the radio jet is oriented towards
us, it is important to validate this assumption. The fact
that the Pa$\alpha $ equivalent
width is noticeably smaller than that for a typical quasar
is consistent with the presence of a beamed continuum
component and may therefore support this assumption. However, this is
not sufficient evidence
in itself. Multi-epoch VLBI imaging of the radio emission could provide
incontrovertible evidence that the radio axis is oriented towards us by
revealing super-luminal motion in the jet.

\section*{ACKNOWLEDGEMENTS}

We thank Chris Tinney and Stuart Ryder of the Anglo-Australian Telescope for
taking the observations and for their assistance during data
reduction. We also thank the referee for his useful comments and
corrections. MJB and JH are supported by PPARC
studentships; MDT is supported by a University of Sheffield
studentship; KAW is supported by a Dorothy Hodgkin Royal Society
Fellowship.. We acknowledge the data analysis facilites at Sheffield
provided
by the Starlink Project, which is run by CCLRC on behalf of PPARC. The
Anglo-Australian Telescope is operated at Siding Springs by the AAO.

\bibliographystyle{mn2e}
\bibliography{abbrev,refs}

\end{document}